\documentstyle[preprint,prd,aps]{revtex}
\begin{document}
\draft
\begin{titlepage}
\preprint{\vbox{\hbox{UDHEP-12-96}
\hbox{hep-th 9701xxx} \hbox{January 1997} }}
\title{ \large \bf Chiral Symmetry Breaking in a Uniform \\
External Magnetic Field} 
\author{\bf D.-S. Lee$^{(a),(b),(c)}$, C. N. Leung$^{(d)}$,
and Y. J. Ng$^{(c)}$}
\address{(a) Institute of Physics, Academia Sinica, Taipei, 
Taiwan\\} 
\address{(b) Department of Physics, Soochow University, Taipei, 
Taiwan\\}
\address{(c) Institute of Field Physics, Department of Physics 
and Astronomy,\\
University of North Carolina, Chapel Hill, NC  27599\\}
\address{(d) Department of Physics and Astronomy, 
University of Delaware\\
Newark, DE 19716 \\}
\maketitle
\begin{abstract}

Using the nonperturbative Schwinger-Dyson equation, we show that 
chiral symmetry is dynamically broken in QED at weak gauge 
couplings when an external magnetic field is present and that chiral 
symmetry is restored at temperatures above $T_c \simeq \frac{\alpha}
{\pi^2} \sqrt{2 \pi |eH|}$, where $\alpha$ is the fine structure 
constant and $H$ is the magnetic field strength.

\end{abstract}
\end{titlepage}

\newpage
\section{Introduction}

Two of us have recently proposed a method to study dynamical chiral 
symmetry breaking in gauge theories in the presence of an external 
field using the Schwinger-Dyson equation approach\cite{LNA}.  It is 
the purpose of this paper to describe the details of the methodology, 
using the case of a uniform magnetic field as an example.  We also 
show how to adopt our formalism to study finite-temperature effects.  

We use quantum electrodynamics (QED) as our model gauge theory and 
consider chiral symmetry breaking in the presence of a constant 
external magnetic field.  We introduce the formalism in Section II 
and derive the Schwinger-Dyson equation for the fermion 
self-energy in the quenched, ladder approximation, Eq.(\ref{SDsimp}).  
Using an approximation suitable for weak gauge couplings, we derive 
in Section III an approximate gap equation, Eq.(\ref{gapeq}), from 
which the infrared dynamical fermion mass, Eq.(\ref{dynamass}), is 
obtained.  Our result is consistent with that found by Gusynin, 
Miransky, and Shovkovy\cite{GMS}, who used a different approach.  
We show in the appendix how our formalism can be applied to the 
approach of Gusynin {\it et al.}, and establish the existence of the 
Nambu-Goldstone boson of the spontaneously broken chiral symmetry.

It has been suggested in the literature\cite{GMS} that the chiral 
symmetry breaking solution may find applications in the electroweak 
phase transition during the early evolution of the universe.  To 
verify this possibility, it is necessary to take into account the 
thermal conditions present in the early universe.  Our formalism 
makes it easier to study such finite-temperature effects.  This is 
discussed in Section IV where we obtain an estimate of the critical 
temperature for chiral symmetry breaking, Eq.(\ref{Tc}).  Our result 
indicates that, in order for the chiral symmetry breaking solution 
found here to be relevant for the electroweak phase transition, an 
unacceptably large magnetic field must be present at the time of 
the phase transition.  We offer our conclusions in Section V.

\newpage
\section{Formalism}

Let us consider chiral symmetry breaking in QED in the presence of 
a static, external electromagnetic field.  The Green's function that 
describes the motion of a fermion with electric charge $e$ in such 
an external field satisfies the equation,
\begin{equation}
\gamma \cdot \Pi(x) G_A(x,y) + \int d^4x' M(x,x') G_A(x',y) = 
\delta^{(4)}(x-y),
\label{Greeneq}
\end{equation}
where $\Pi_\mu(x) = - i \partial_\mu - e A_\mu^{\rm ext}(x)$, and 
$M(x,x')$ is the mass operator $\hat{M}$ in the coordinate 
representation: $M(x,x')~=~\langle x|\hat{M}|x'\rangle$.  As 
pointed out by Ritus\cite{Ritus}, $\hat{M}$ is a scalar 
$\gamma$-matrix function of the $\Pi_\mu$ and the $F_{\mu \nu}~=~
\partial_\mu A_\nu^{\rm ext} - \partial_\nu A_\mu^{\rm ext}$, 
and for constant $F_{\mu \nu}$, 
\begin{equation}
\hat{M} = \hat{M}(\gamma^\mu \Pi_\mu, \sigma^{\mu \nu} F_{\mu \nu}, 
(F_{\mu \nu} \Pi^\nu)^2, \gamma_5 F_{\mu \nu} \tilde{F}^{\mu \nu}).
\label{mop}
\end{equation}
In other words, for uniform external fields, only four independent 
$\gamma$-matrix valued scalars can be formed out of $\Pi_\mu$ and 
$F_{\mu \nu}$, as listed in Eq.(\ref{mop}), where $\tilde{F}^{\mu \nu} 
\equiv \frac{1}{2} \varepsilon^{\mu \nu \lambda \tau} F_{\lambda \tau}$.  
Furthermore, all these four scalars commute with $(\gamma \cdot \Pi)^2$; 
consequently,
\begin{equation}
[\hat{M}, (\gamma \cdot \Pi)^2] = 0.
\label{crM}
\end{equation}
It follows that the mass operator will be diagonal in the basis 
spanned by the eigenfunctions of $(\gamma \cdot \Pi)^2$:
\begin{equation}
- (\gamma \cdot \Pi)^2  \psi_p(x) = p^2 \psi_p(x).
\label{eigeneq}
\end{equation}
If we work in the chiral representation in which $\Sigma_3~=~i 
\gamma_1 \gamma_2$ and $\gamma_5$ are both diagonal with 
eigenvalues $\sigma = \pm 1$ and $\chi = \pm 1$, respectively, 
the eigenfunctions of $(\gamma \cdot \Pi)^2$ has the general form
\begin{equation}
\psi_p(x) = E_{p\sigma\chi}(x) \omega_{\sigma\chi},
\label{eigenfcn0}
\end{equation}
where $\omega_{\sigma\chi}$ are bispinors which are the eigenvectors 
of $\Sigma_3$ and $\gamma_5$.  The exact functional form of the 
$E_{p\sigma\chi}(x)$ will depend on the specific external field 
configuration.  Our method is based on the use of these 
eigenfunctions as basis functions.  This is a natural choice as they 
are the wavefunctions of the asymptotic states when a uniform external 
field is present.  The advantage of using this representation 
is obvious: $\hat{M}$ can now be put in terms of its eigenvalues, so 
the problems arising from its dependence on the operator $\Pi$ can be 
avoided.

We now restrict our consideration to the case of a constant magnetic 
field of strength $H$, the vector potential of which may be taken to 
be $A_\mu^{\rm ext} = (0, 0, Hx_1, 0)$, $\mu = 0, 1, 2, 3$.  Our 
metric has the signature $g_{\mu \nu} = {\rm diag}(-1, 1, 1, 1)$.  
The eigenfunctions $E_{p\sigma\chi}(x)$ are now given by
\begin{equation}
E_{p\sigma}(x) = N {\rm e}^{i (p_0x^0 + p_2x^2 + p_3x^3)} 
D_n(\rho), 
\label{eigenfcn}
\end{equation}
where $N$ is a normalization factor and $D_n(\rho)$ are the 
parabolic cylinder functions\cite{math} with argument 
$\rho = \sqrt{2 |e H|} (x_1 - \frac{p_2}{e H})$ and indices 
(which are the quantum numbers of the Landau levels)
\begin{equation}
n = n(k,\sigma) \equiv k + \frac{e H \sigma}{2 |e H|} - \frac{1}{2},
~~~~n = 0, 1, 2, ...
\label{index}
\end{equation}         
Note that, in the absence of an external electric field, the 
eigenfunctions do not depend on $\chi$.  The eigenvalue $p$ 
stands for the set $(p_0, p_2, p_3, k)$, where $k$ is the 
quantum number of the quantized squared transverse momentum:
\begin{eqnarray}
- (\gamma \cdot \Pi_\perp)^2  \psi_p(x) && ~\equiv 
- (\gamma^1 \Pi_1 + \gamma^2 \Pi_2)^2 \psi_p(x) 
\nonumber\\
~&& ~= (\Pi_1^2 + \Pi_2^2 - e H \Sigma_3)\psi_p(x) 
\nonumber\\
~&& ~= p_\perp^2 \psi_p(x) \nonumber \\
~&& ~\equiv 2 |eH| k \psi_p(x).
\label{pperp}
\end{eqnarray}
Note that $(\gamma \cdot \Pi)^2~=~(\gamma \cdot \Pi_\parallel)^2 + 
(\gamma \cdot \Pi_\perp)^2$, where $(\gamma \cdot \Pi_\parallel)^2~
\equiv~(\gamma^0 \Pi_0 + \gamma^3 \Pi_3)^2~=~\Pi_0^2 - \Pi_3^2$, 
hence $p^2 = - p_0^2 + p_3^2 + 2 |eH| k$.  The allowed values for 
$k$ are seen from Eq.(\ref{index}) to be $k = 0,~1,~2,~...~$.

Following Ritus\cite{Ritus}, we form the eigenfunction-matrices 
$E_p(x)$ = diag($E_{p11}(x)$, $E_{p-11}(x)$, $E_{p1-1}(x)$, 
$E_{p-1-1}(x)$).  As noted above, in a pure magnetic field, 
references to $\chi$ are irrelevant and can be dropped, hence 
\begin{eqnarray}
E_p(x) ~&&~=~\sum_\sigma E_{p\sigma}(x) {\rm diag} 
(\delta_{\sigma 1}, \delta_{\sigma -1}, \delta_{\sigma 1}, 
\delta_{\sigma -1}) \nonumber\\
~&&~\equiv~\sum_\sigma E_{p\sigma}(x) \Delta(\sigma).
\label{ep}
\end{eqnarray}
Using the orthogonal property of the parabolic cylinder functions
\cite{WW},
\begin{equation}
\int_{- \infty}^\infty d\rho D_{n'}(\rho) D_n(\rho) = \sqrt{2 \pi} 
n! \delta_{nn'},
\end{equation}
it is straightforward to establish that the $E_p$ are orthonormal 
$(\bar{E}_p \equiv \gamma^0 E_p^\dagger \gamma^0)$:
\begin{eqnarray}
\int d^4x \bar{E}_{p'}(x) E_p(x) && ~= (2 \pi)^4 \hat{\delta}^{(4)}
(p-p')
\nonumber \\
~&& ~\equiv (2 \pi)^4 \delta_{kk'} \delta(p_0 - p'_0) 
\delta(p_2 - p'_2) \delta(p_3 - p'_3)
\label{orthonormal}
\end{eqnarray}
as well as complete:
\begin{equation}
\Sigma \!\!\!\!\! \displaystyle{\int} d^4p E_p(x) \bar{E}_p(y) 
= (2 \pi)^4 \delta^{(4)}(x-y),~~~~~~\Sigma \!\!\!\!\!\! \int d^4p 
\equiv \sum_{k} \int dp_0 dp_2 dp_3
\label{complete}
\end{equation}
provided that the normalization constant in Eq.(\ref{eigenfcn}) 
is taken to be $N(n) = (4 \pi |eH|)^{1/4}/\sqrt{n!}$.  Since 
the $E_p$ are linear combinations of the eigenfunctions of the 
mass operator, they satisfy 
\begin{equation}
\int d^4x' M(x,x') E_p(x') = E_p(x) \tilde{\Sigma}_A(\bar{p}),
\label{masseigeneq}
\end{equation}
where $\tilde{\Sigma}_A(\bar{p})$ represents the eigenvalue matrix 
of the mass operator.  The $E_p$ also satisfy the important property 
that 
\begin{equation}
\gamma \cdot \Pi~E_p(x) = E_p(x)~\gamma \cdot \bar{p},
\label{PROPERTY}
\end{equation}
where $\bar{p}_0 = p_0,~\bar{p}_1 = 0,~\bar{p}_2 
= - {\rm sgn}(eH) \sqrt{2|eH|k},~\bar{p}_3 = p_3$.  Note that, 
due to the rotational symmetry about the direction of the 
magnetic field, the system is effectively a (2+1)-dimensional 
one, as is evident from the momentum $\bar{p}$.

By using the above properties of the $E_p$-functions, the fermion 
Green's function may be expressed as 
\begin{equation}
G_A(x,y) = \Sigma \!\!\!\!\!\! \int \frac{d^4p}{(2 \pi)^4} E_p(x) 
\frac{1}{\gamma \cdot \bar{p} + \tilde{\Sigma}_A(\bar{p})} 
\bar{E}_p(y),
\label{Greenfcn}
\end{equation}
Eqs.(\ref{PROPERTY}) and (\ref{masseigeneq}) guarantee that 
Eq.(\ref{Greeneq}) is satisfied.  It follows that, in the 
$E_p$-representation,
\begin{eqnarray}
G_A(p,p')~&&~\equiv \int d^4x d^4y \bar{E}_p(x) G_A(x,y) 
E_{p'}(y) \nonumber\\
~&&~= (2 \pi)^4 \hat{\delta}^{(4)}(p-p') \frac{1}{\gamma 
\cdot \bar{p} + \tilde{\Sigma}_A(\bar{p})}
\label{pGreenfcn}
\end{eqnarray}
which shows explicitly that the fermion propagator is diagonal 
(in momentum) in this representation.  Similarly, the mass 
operator may be written in the $E_p$-representation as 
\begin{eqnarray}
M(p,p') ~&&~= \int d^4x d^4x' \bar{E}_p(x) M(x,x') E_{p'}(x')
\nonumber\\ 
~&&~= (2 \pi)^4 \hat{\delta}^{(4)}(p-p') \tilde
{\Sigma}_A(\bar{p}).
\label{pmassop}
\end{eqnarray}

We may now write down the Schwinger-Dyson (SD) equation for 
the fermion self-energy.  We shall work in the quenched, 
ladder approximation in which 
\begin{equation}
M(x,x') = i e^2 \gamma^\mu G_A(x,x') \gamma^\nu D_{\mu\nu}
(x-x'),
\label{massop}
\end{equation}
where $D_{\mu\nu}(x-x')$ is the bare photon propagator,
\begin{equation}
D_{\mu\nu}(x-x') = \int \frac{d^4q}{(2 \pi)^4} \frac{{\rm e}^
{i q\cdot(x-x')}}{q^2 - i\epsilon} \left(g_{\mu\nu} -(1 - \xi) 
\frac{q_\mu q_\nu}{q^2}\right),
\label{photon}
\end{equation}
The SD equation in this approximation reads
\begin{eqnarray}
(2 \pi)^4 \hat{\delta}^{(4)}(p-p') \tilde{\Sigma}_A(\bar{p}) 
&=& 
i e^2 \int d^4x d^4x' \Sigma \!\!\!\!\!\! \int \frac{d^4p''}
{(2 \pi)^4} \bar{E}_p(x) \gamma^\mu E_{p''}(x) \nonumber \\
& &
\cdot~ \frac{1}{\gamma 
\cdot \bar{p}'' + \tilde{\Sigma}_A(\bar{p}'')} ~\cdot~ 
\bar{E}_{p''}(x') \gamma^\nu E_{p'}(x') D_{\mu\nu}(x-x'). 
\label{SDfull}
\end{eqnarray}

The SD Eq.(\ref{SDfull}) can be simplified by performing 
the integrations over $x$, $x'$, $p''_0$, $p''_2$, and 
$p''_3$ exactly.  Consider first the $x$-integrals.  The 
$x_0$-, $x_2$-, and $x_3$-integration each yields a 
$\delta$-function, leaving 
\begin{eqnarray}
\int d^4x \bar{E}_p(x) \gamma^\mu E_{p''}(x) {\rm e}^
{iq \cdot x}
&~=~& 
(2 \pi)^3 \delta^{(3)}(p''+q-p) \sum_{\sigma,\sigma''} N(n) 
N(n'') \nonumber \\
& &
\cdot \int_{-\infty}^\infty dx_1 
D_n(\rho) D_{n''}(\rho'') {\rm e}^{iq_1x_1} \gamma^0 \Delta 
\gamma^0 \gamma^\mu \Delta'',
\label{xint}
\end{eqnarray}
where $\delta^{(3)}(p''+q-p) \equiv \delta (p_0''+q_0-p_0) 
\delta (p_2''+q_2-p_2) \delta (p_3''+q_3-p_3)$, 
$\rho'' = \sqrt{2 |e H|} (x_1 - \frac{p_2''}{e H})$, 
$n'' = n(k'',\sigma'')$, and $\Delta'' = \Delta(\sigma'')$.
Due to the presence of the $\delta (p_2''+q_2-p_2)$, the 
remaining $x_1$-integral may be more conveniently written as 
\begin{equation}
\int_{-\infty}^\infty dx_1 {\rm e}^{iq_1x_1} D_n(\rho)  
D_{n''}(\rho'') = \frac{1}{\sqrt{2 |eH|}} {\rm e}^{i q_1 
(p_2''+p_2)/(2eH)} I_{nn''}(\hat{q}_1,\hat{q}_2), 
\label{x1int}
\end{equation}
where 
\begin{equation}
I_{nn''}(\hat{q}_1,\hat{q}_2) \equiv \int_{-\infty}^\infty 
d\eta {\rm e}^{i \eta {\rm sgn}(eH) \hat{q}_1} D_n(\eta - 
\hat{q}_2) D_{n''}(\eta + \hat{q}_2),
\label{Idefn}
\end{equation}
$\eta~\equiv~\rho + \hat{q}_2$, and $\hat{q}_\mu$ 
are dimensionless variables defined as 
\begin{equation}
\hat{q}_\mu \equiv \frac{q_\mu \sqrt{2 |eH|}}{2eH},~~~~~~~~~~
\mu = 0,~1,~2,~3.
\label{qhat}
\end{equation}

If we transform to the polar coordinates: ($\hat{q}_\perp 
\equiv \sqrt{\hat{q}_1^2 + \hat{q}_2^2}$, $\varphi 
\equiv {\rm arctan}(\hat{q}_2/\hat{q}_1)$), we can evaluate
the $I_{nn''}$ by first noting that they satisfy\cite{Ritus} 
\begin{equation}
\frac{\partial I_{nn''}(\hat{q}_\perp, \varphi)}{\partial 
\varphi} = i {\rm sgn}(eH) (n - n'') I_{nn''}(\hat{q}_\perp, 
\varphi).
\end{equation}
Hence, 
\begin{equation}
I_{nn''}(\hat{q}_\perp, \varphi) = I_{nn''}(\hat{q}_\perp) 
{\rm e}^{i {\rm sgn}(eH) (n - n'') \varphi},
\end{equation}
where 
\begin{equation}
I_{nn''}(\hat{q}_\perp) \equiv \int_{-\infty}^\infty d\eta 
{\rm e}^{i \eta {\rm sgn}(eH) \hat{q}_\perp} D_n(\eta) 
D_{n''}(\eta).
\label{Iqperp}
\end{equation}
Note that $I_{nn''}(\hat{q}_\perp) = I_{n''n}(\hat{q}_\perp)$.
To compute $I_{nn''}(\hat{q}_\perp)$, we use the relation,

\begin{equation}
D_n(\eta) D_{n''}(\eta) = {\rm e}^{- \eta^2/4} \sum_{m=0}^
{{\rm min}(n, n'')} \frac{n! n''!}{m! (n-m)! (n''-m)!} 
D_{n+n''-2m}(\eta),
\end{equation}
and the Rodrigues formula,

\begin{equation}
D_n(\eta) = (-1)^n {\rm e}^{\eta^2/4} \frac{{\rm d}^n}
{{\rm d}\eta^n} {\rm e}^{- \eta^2/2},
\end{equation}
to secure

\begin{equation}
I_{nn''}(\hat{q}_\perp) = \sqrt{2 \pi} {\rm e}^{-\hat{q}_
\perp^2/2} J_{nn''}(\hat{q}_\perp),
\label{Ifinal}
\end{equation}
where

\begin{equation}
J_{nn''}(\hat{q}_\perp) \equiv \sum_{m=0}^{{\rm min}(n, n'')} 
\frac{n! n''!}{m! (n-m)! (n''-m)!} [i {\rm sgn}(eH) 
\hat{q}_\perp]^{n+n''-2m}.
\label{defJ}
\end{equation}

Putting all the pieces together, we have
\begin{eqnarray}
\int d^4x \bar{E}_p(x) \gamma^\mu E_{p''}(x) {\rm e}^{iq 
\cdot x}
&~=~& 
(2 \pi)^4 \delta^{(3)}(p''+q-p) {\rm e}^{i q_1 (p_2''+p_2)
/(2eH)} \nonumber \\
& &
\!\!\!\!\!\!\!\!\!\!\!\!
\cdot~ {\rm e}^{-\hat{q}_\perp^2/2} \sum_{\sigma,\sigma''} 
\frac{1}{\sqrt{n! n''!}} {\rm e}^{i {\rm sgn}(eH) (n - n'') 
\varphi} J_{nn''}(\hat{q}_\perp) \Delta \gamma^\mu \Delta'',
\label{xintfinal}
\end{eqnarray}
where we have used the identity, $\gamma^0 \Delta \gamma^0 = 
\Delta$.  Similarly, the $x'$-integrals yield
\begin{eqnarray}
\int d^4x' \bar{E}_{p''}(x') \gamma^\nu E_{p'}(x') 
{\rm e}^{-iq \cdot x'}
&~=~& 
(2 \pi)^4 \delta^{(3)}(p''+q-p') {\rm e}^{-i q_1 (p_2''+p_2')
/(2eH)} \nonumber \\
& &
\!\!\!\!\!\!\!\!\!\!\!\!
\cdot~ {\rm e}^{-\hat{q}_\perp^2/2} \sum_{\sigma',\tilde
{\sigma}''} \frac{1}{\sqrt{n'! \tilde{n}''!}} {\rm e}^{i 
{\rm sgn}(eH) (\tilde{n}'' - n') \varphi} J_{\tilde{n}''n'}
(\hat{q}_\perp) \tilde{\Delta}'' \gamma^\nu \Delta',
\label{x'intfinal}
\end{eqnarray}
where $n' = n(k',\sigma')$, $\tilde{n}'' = n(k'',\tilde
{\sigma}'')$, $\Delta' = \Delta(\sigma')$, and 
$\tilde{\Delta}'' = \Delta(\tilde{\sigma}'')$.  The presence 
of the $\delta$-functions in Eqs.(\ref{xintfinal}) and 
(\ref{x'intfinal}) allows easy integrations over $p''_0$, 
$p''_2$, and $p''_3$ in Eq.(\ref{SDfull}), yielding 
$\delta^{(3)}(p-p') = \delta(p_0-p_0') \delta (p_2-p_2') 
\delta (p_3-p_3')$ which matches that on the left hand side 
of Eq.(\ref{SDfull}).  The SD equation is therefore reduced 
to   
\begin{eqnarray}
\tilde{\Sigma}_A(\bar{p}) \delta_{kk'} 
&~=~&
i e^2 (2 |eH|) \sum_{k''} \sum_{\{\sigma\}} \int \frac
{d^4\hat{q}}{(2 \pi)^4} 
\frac{{\rm e}^{i {\rm sgn}(eH)(n-n''+\tilde{n}''-n')\varphi}}
{\sqrt{n!n'!n''!\tilde{n}''!}} 
\nonumber \\
& &
\cdot~ {\rm e}^{- \hat{q}_\perp^2} J_{nn''}(\hat{q}_\perp) 
J_{\tilde{n}''n'}(\hat{q}_\perp) \frac{1}{\hat{q}^2} 
\left(g_{\mu\nu} - (1 - \xi) \frac{\hat{q}_\mu \hat{q}_\nu}
{\hat{q}^2}\right) \nonumber \\
& &
\cdot~ \Delta \gamma^\mu \Delta'' \frac{1}{\gamma \cdot 
\bar{p}'' + \tilde{\Sigma}_A(\bar{p}'')} \tilde{\Delta}'' 
\gamma^\nu \Delta',
\label{SDsimp}
\end{eqnarray}
where the summation over $\{\sigma\}$ means summing over 
$\sigma$, $\sigma'$, $\sigma''$, and $\tilde{\sigma}''$, 
and the momentum $\bar{p}''$ is understood to be: 
$\bar{p}''_0 = p_0 - q_0$, $\bar{p}''_1 = 0$, 
$\bar{p}''_2 = -~{\rm sgn}(eH) \sqrt{2|eH|k''}$, 
$\bar{p}''_3 = p_3 - q_3$.

\newpage
\section{Solution to the Schwinger-Dyson Equation}

An approximate solution to Eq.(\ref{SDsimp}) may be obtained 
by observing that, due to the factor e$^{- \hat{q}_\perp^2}$ 
in the integrand, contributions from large values of 
$\hat{q}_\perp$ are suppressed.  Thus, by keeping only the 
terms with the smallest power of $\hat{q}_\perp$ in 
$J_{nn''}(\hat{q}_\perp)$, i.e., 

\begin{eqnarray}
J_{nn''}(\hat{q}_\perp) &\rightarrow& \frac{[{\rm max}(n, n'')]!}
{|n-n''|!} (i {\rm sgn}(eH) \hat{q}_\perp)^{|n-n''|}\nonumber\\
&\rightarrow& n!~\delta_{nn''}
\label{Japprox}
\end{eqnarray}
and similarly for $J_{\tilde{n}''n'}(\hat{q}_\perp)$, the SD 
equation is simplified to 

\begin{eqnarray}
\tilde{\Sigma}_A(\bar{p}) \delta_{kk'} 
&~\simeq~&
i e^2 (2 |eH|) \sum_{k''} \sum_{\{\sigma\}} \delta_{nn''} 
\delta_{\tilde{n}''n'} \int \frac{d^4\hat{q}}{(2 \pi)^4} 
\frac{{\rm e}^{- \hat{q}_\perp^2}}{\hat{q}^2} 
\left(g_{\mu\nu} - (1 - \xi) \frac{\hat{q}_\mu \hat{q}_\nu}
{\hat{q}^2}\right) 
\nonumber \\
& &
\cdot~ \Delta \gamma^\mu \Delta'' \frac{1}{\gamma \cdot 
\bar{p}'' + \tilde{\Sigma}_A(\bar{p}'')} \tilde{\Delta}'' 
\gamma^\nu \Delta'.
\label{SDapprox}
\end{eqnarray}
The solution is expected to have the form $\tilde{\Sigma}_
A(\bar{p}) = Z(\bar{p}) \gamma \cdot \bar{p} + 
\Sigma_A(\bar{p})$, where $\Sigma_A(\bar{p})$ is the 
dynamically generated fermion mass and is assumed to be 
proportional to the unit matrix (see remarks after 
Eq.(\ref{SDapprox3})).  Eq.(\ref{SDapprox}) then reads 

\begin{eqnarray}
\left[Z(\bar{p}) \gamma \cdot \bar{p} + \Sigma_A(\bar{p})
\right] \delta_{kk'} 
&~\simeq~&
i e^2 (2 |eH|) \sum_{k''} \sum_{\{\sigma\}} \delta_{nn''} 
\delta_{\tilde{n}''n'} \int \frac{d^4\hat{q}}{(2 \pi)^4} 
\frac{{\rm e}^{- \hat{q}_\perp^2}}{\hat{q}^2} \nonumber \\ 
& &
\!\!\!\!\!\!\!\!\!\!\!\!\! \cdot~ \frac{\Sigma_A(\bar{p}'') 
[G_1 - (1 - \xi) W_1] - [1 + Z(\bar{p}'')] 
[G_2 - (1 - \xi) W_2]}
{[1+Z(\bar{p}'')]^2 \bar{p}''^2 + \Sigma_A^2(\bar{p}'')}, 
\label{SDapprox2}
\end{eqnarray}
where 

\begin{eqnarray}
G_1 & ~\equiv~ &
\Delta \gamma^\mu \Delta'' \tilde{\Delta}'' 
\gamma_\mu \Delta', \nonumber \\
W_1 & ~\equiv~ &
\frac{1}{\hat{q}^2} \Delta (\gamma \cdot \hat{q}) \Delta'' 
\tilde{\Delta}'' (\gamma \cdot \hat{q}) \Delta', \nonumber\\
G_2 & ~\equiv~ & 
\Delta \gamma^\mu \Delta'' (\gamma \cdot \bar{p}'') 
\tilde{\Delta}'' \gamma_\mu \Delta', \nonumber \\
W_2 & ~\equiv~ & 
\frac{1}{\hat{q}^2} \Delta (\gamma \cdot \hat{q}) 
\Delta'' (\gamma \cdot \bar{p}'') \tilde{\Delta}'' 
(\gamma \cdot \hat{q}) \Delta'.
\label{GWdef}
\end{eqnarray}

The matrices $G_{1,2}$ and $W_{1,2}$ may be simplified as 
follows.  First we note that the $\Delta$-matrices may be 
expressed as

\begin{equation}
\Delta(\sigma) = \frac{1}{2} \left(1 + D_\sigma \Sigma_3 
\right),
\end{equation}
where $D_\sigma \equiv (\delta_{\sigma 1} - \delta_{\sigma -1})$, 
$D^2 = 1$.  They also satisfy the commutation relations, 

\begin{equation}
\Delta \gamma^\mu_\perp = \gamma^\mu_\perp \nabla
\end{equation}
and
\begin{equation}
[\Delta, \gamma^\mu_\parallel] = 0 = 
[\nabla, \gamma^\mu_\parallel],
\end{equation}
where the subscript $\perp$ refers to the transverse components, 
$\mu = 1, 2$, the subscript $\parallel$ refers to the 
longitudinal components, $\mu = 0, 3$, and $\nabla$ is the 
complement of $\Delta$:

\begin{equation}
\nabla(\sigma) = 1 - \Delta(\sigma) = \frac{1}{2} 
\left(1 - D_\sigma \Sigma_3 \right).
\end{equation}
Secondly, products of the $\Delta$-matrices may be expressed 
in terms of a single $\Delta$-matrix, e.g., 

\begin{equation}
\Delta(\sigma) \Delta(\sigma'') = \delta_{\sigma \sigma''} 
\Delta(\sigma),
\end{equation}
and similarly for the $\nabla$-matrices.  Using these relations, 
we find that, after performing the summation over the spin 
indices, 

\begin{eqnarray}
\sum_{\{\sigma\}} \delta_{nn''} \delta_{\tilde{n}''n'} G_1 
&~=~& - 2 \delta_{kk'} \left[\delta_{kk''} + 
\delta_{k, k''-{\rm sgn}(eH)} \Delta(1) + 
\delta_{k, k''+{\rm sgn}(eH)} \Delta(-1) \right], \\
\label{G1summed}
\sum_{\{\sigma\}} \delta_{nn''} \delta_{\tilde{n}''n'} W_1 
&~\simeq~& - \delta_{kk'} \delta_{kk''}, \\
\label{W1summed}
\sum_{\{\sigma\}} \delta_{nn''} \delta_{\tilde{n}''n'} G_2 
&~=~& 2 \delta_{kk'} \left\{\delta_{kk''} \left(\gamma \cdot 
\bar{p}_\perp\right) + \gamma \cdot \left(\bar{p}_\parallel - 
q_\parallel \right) \cdot \right. \nonumber \\
& &
~~~~~~\left. \left[\delta_{k, k''-{\rm sgn}(eH)} \Delta(1) + 
\delta_{k, k''+{\rm sgn}(eH)} \Delta(-1)\right] \right\}, \\
\label{G2summed}
\sum_{\{\sigma\}} \delta_{nn''} \delta_{\tilde{n}''n'} W_2 
&~\simeq~& \delta_{kk'} \delta_{kk''} \left\{\gamma \cdot 
\bar{p}_\perp + \frac{1}{\hat{q}^2} 
\left(\gamma \cdot \hat{q}_\parallel \right) 
\left[\gamma \cdot (\bar{p}_\parallel - q_\parallel) \right] 
\left(\gamma \cdot \hat{q}_\parallel \right)\right\}.
\label{W2summed}
\end{eqnarray}
where $\gamma \cdot \bar{p}_\perp$ = $\gamma^2 \bar{p}_2$, 
$\Delta(1)$ = diag(1, 0, 1, 0), and $\Delta(-1)$ = 
diag(0, 1, 0, 1).  In accordance with the small $\hat{q}_\perp$ 
approximation used on the $J_{nn''}$ and $J_{\tilde{n}''n'}$, 
Eq.(\ref{Japprox}), terms in $W_1$ and $W_2$ that are proportional 
to $\hat{q}_\perp$ have been dropped.  It is satisfying that the 
spin summation produces the Kronecker delta, $\delta_{kk'}$, which 
matches the one on the left hand side of Eq.(\ref{SDapprox2}).  
Note also that, due to the restriction $n = n''$ which arises from 
the small $\hat{q}_\perp$ approximation, the summation over $k''$ 
is now restricted to only three terms: for a given $k$, 
$k'' = k,~k \pm 1$.  

The SD equation may now be written as 
\begin{eqnarray}
& & Z(\bar{p}) \gamma \cdot \bar{p} + \Sigma_A(\bar{p})
\nonumber \\
&~\simeq~&
-~i e^2 (2 |eH|) \sum_{k''} \int \frac{d^4\hat{q}}{(2 \pi)^4} 
\frac{{\rm e}^{- \hat{q}_\perp^2}}{\hat{q}^2} \frac{1}
{[1+Z(\bar{p}'')]^2 \bar{p}''^2 + \Sigma_A^2(\bar{p}'')} 
\nonumber \\ 
& &
\cdot~\left\{\left[1 + Z(\bar{p}'')\right] 
\left[\delta_{k''k} \left((1 + \xi)\gamma \cdot \bar{p}_\perp 
- \frac{1-\xi}{\hat{q}^2} 
\left(\gamma \cdot \hat{q}_\parallel \right)
\gamma \cdot \left(\bar{p}_\parallel - q_\parallel \right)
\left(\gamma \cdot \hat{q}_\parallel \right) \right)\right.\right. 
\nonumber \\
& &
\left.\left.~~~~~~~~~+ 2\left(\delta_{k'', k+{\rm sgn}(eH)} \Delta(1) 
+ \delta_{k'', k-{\rm sgn}(eH)} \Delta(-1)\right) \gamma \cdot 
\left(\bar{p}_\parallel - q_\parallel \right) \right] \right.
\nonumber \\
& &
\left. ~~~+ \Sigma_A(\bar{p}'') \left[\delta_{k''k} (1 + \xi) 
+ 2\left(\delta_{k'', k+{\rm sgn}(eH)} \Delta(1) + 
\delta_{k'', k-{\rm sgn}(eH)} \Delta(-1)\right)\right] \right\}. 
\label{SDapprox3}
\end{eqnarray}
Recall that $\bar{p}''^2~=~(\bar{p}_\parallel - q_\parallel)^2 + 
2|eH|k''$.  

Eq.(\ref{SDapprox3}) shows that our earlier assumption of 
$\Sigma_A(\bar{p})$ being proportional to the unit matrix is 
correct only for the $k'' = k$ term.  The reason is that $\gamma 
\cdot \Pi$, and hence the mass operator $\hat{M}$, 
does not commute with $\Sigma_3$.  However, if we consider only 
the low energy ($\bar{p}^2 \ll |eH|$) behaviors, in particular, 
in the $\bar{p}_\perp$ = 0 = $k$ limit, the $k'' = 0$ term will 
dominate and we obtain the approximate SD equation,
\begin{eqnarray}
& & Z(\bar{p}_\parallel) \gamma \cdot \bar{p}_\parallel + 
\Sigma_A(\bar{p}_\parallel) \nonumber \\
&~\simeq~&
-~i e^2 (2 |eH|) \int \frac{d^4\hat{q}}{(2 \pi)^4} 
\frac{{\rm e}^{- \hat{q}_\perp^2}}{\hat{q}^2} \frac{1}
{[1+Z(\bar{p}_\parallel-q_\parallel)]^2 (\bar{p}_\parallel - 
q_\parallel)^2 + \Sigma_A^2(\bar{p}_\parallel-q_\parallel)} 
\nonumber \\ 
& &
\cdot~\left\{\left[1 + Z(\bar{p}_\parallel-q_\parallel)\right] 
\frac{\xi-1}{\hat{q}^2} 
\left(\gamma \cdot \hat{q}_\parallel \right)
\gamma \cdot \left(\bar{p}_\parallel - q_\parallel \right)
\left(\gamma \cdot \hat{q}_\parallel \right) 
+ \Sigma_A(\bar{p}_\parallel-q_\parallel) (1 + \xi)
\right\}.
\label{SDparallel}
\end{eqnarray}
In this case $\Sigma_A$ is proportional to the unit matrix.  
This approximation is equivalent to the lowest Landau level 
approximation employed in \cite{GMS}.

We see from Eq.(\ref{SDparallel}) that, in the Feynman gauge 
($\xi = 1$), $Z(\bar{p}_\parallel) = 0$ and the SD equation 
for the dynamically generated fermion mass becomes 
\begin{equation}
\Sigma_A(\bar{p}_\parallel) ~\simeq~
e^2 (4 |eH|) \int \frac{d^4\hat{q}}{(2 \pi)^4} 
\frac{{\rm e}^{- \hat{q}_\perp^2}}{\hat{q}^2} 
\frac{\Sigma_A(\bar{p}_\parallel-q_\parallel)}
{(\bar{p}_\parallel - q_\parallel)^2 
+ \Sigma_A^2(\bar{p}_\parallel-q_\parallel)}, 
\label{SDsigma}
\end{equation}
where we have made a Wick rotation to Euclidean space: $p_0 
\rightarrow ip_4$, $q_0 \rightarrow iq_4$.  
This can be turned into a differential equation for $\Sigma_A$,
as was done in Ref. \cite{Hong} where the momentum dependence of 
the dynamical mass is discussed.  Here we content ourselves with 
finding a solution for the infrared fermion mass scale,
\begin{equation}
\Sigma_A(0) ~\simeq~
e^2 (4 |eH|) \int \frac{d^4\hat{q}}{(2 \pi)^4} 
\frac{{\rm e}^{- \hat{q}_\perp^2}}{\hat{q}^2} 
\frac{\Sigma_A(q_\parallel)}{2 |eH| \hat{q}_\parallel^2 
+ \Sigma_A^2(q_\parallel)}. 
\label{SDsigma0}
\end{equation}
Since $\Sigma_A(q_\parallel)$ is expected to diminish with 
increasing $q_\parallel^2$, the integral in Eq.(\ref{SDsigma0}) 
is dominated by the contributions from small $\hat{q}_\parallel^2$.  
It is therefore reasonable to approximate $\Sigma_A(q_\parallel)$ 
in the integrand by $\Sigma_A(0)$ = $m \times {\bf 1}$, where $m$ 
is the dynamical mass and {\bf 1} is the unit matrix, thus 
securing the gap equation,
\begin{eqnarray}
1 &~\simeq~& e^2 (4 |eH|) \int \frac{d^4\hat{q}}{(2 \pi)^4} 
\frac{{\rm e}^{- \hat{q}_\perp^2}}{\hat{q}^2} 
\frac{1}{2 |eH| \hat{q}_\parallel^2 + m^2} \nonumber \\
&~\simeq~& \frac{e^2}{4 \pi^2} |eH| \int_0^\infty 
d\hat{q}_\perp^2 \int_0^\infty d\hat{q}_\parallel^2 
\frac{{\rm e}^{- \hat{q}_\perp^2}}
{\hat{q}_\perp^2 + \hat{q}_\parallel^2} 
\frac{1}{2 |eH| \hat{q}_\parallel^2 + m^2} \nonumber \\
&~\simeq~& \frac{\alpha}{\pi} |eH| \int_0^\infty d\hat{q}_\perp^2 
\frac{{\rm e}^{- \hat{q}_\perp^2} \ln(2|eH| \hat{q}_\perp^2/m^2)}
{2|eH| \hat{q}_\perp^2 - m^2}. 
\label{gapeq}
\end{eqnarray}

The solution to Eq.(\ref{gapeq}) has the form 
\begin{equation}
m \simeq a~\sqrt{|eH|}~{\rm e}^{- b\sqrt{\pi/\alpha}},
\label{dynamass}
\end{equation} 
where $a$ and $b$ are positive constants of order 1.  The  
nonperturbative nature of this result is apparent.  Furthermore, 
according to the last equality of Eq.(\ref{gapeq}), the 
dominant contributions to the integral come from the region 
$2|eH| \hat{q}_\perp^2 \sim m^2$.  Consistency with our small 
$\hat{q}_\perp$ assumption requires that $m \ll \sqrt{|eH|}$, 
which in turn requires that $\alpha \ll 1$.  In other words, 
the  solution for the dynamical mass found above applies to the 
weak-coupling regime of QED.  

To establish that the above solution to the SD equation for the 
fermion self-energy does indeed correspond to a dynamical chiral 
symmetry breaking solution, it is necessary to demonstrate the 
existence of the corresponding Nambu-Goldstone (NG) boson.  One 
way to establish this is by studying the Bethe-Salpeter equation 
of the bound-state NG boson, as was done by Gusynin {\it et al.}
\cite{GMS}, who found a solution consistent with our 
Eq.(\ref{dynamass}).  We show in the Appendix how the same solution, 
Eq.(\ref{dynamass}), can be obtained from the Bethe-Salpeter 
equation for the NG boson, using the $E_p$-representation of the 
fermion progagator.  This helps to justify that an external magnetic 
field serves as a catalyst for chiral symmetry breaking in QED.

\newpage
\section{Symmetry Breaking at Nonzero Temperature}

The formalism described above is very useful for studying 
dynamical symmetry breaking in an external field at nonzero 
temperatures.  As expected on physical grounds\cite{Ng}, the 
long-range order of a system decreases as the temperature 
increases and chiral symmetry will generally be restored 
above a certain critical temperature.  We shall see that 
this expectation is indeed correct for the case of chiral 
symmetry breaking in a magnetic field and an estimate of 
the critical temperature will be obtained.

To incorporate the thermal effects, we use the imaginary 
time and energy formalism \cite{DJ} in which
\begin{equation}
    0 \le i x_0 \le \beta
\end{equation}
and 
\begin{eqnarray}
  && q_0 = 2l'\pi / (-i \beta)  ~~{\rm{for ~ bosons}}, 
~~ l' =0, \pm 1, \pm 2,... \nonumber \\
  && p_0 = (2l+1)\pi / (-i \beta) ~~ {\rm{for ~ fermions}}, 
~~ l=0, \pm 1, \pm 2,... 
\end{eqnarray}
where $\beta  = 1/T$, with the Boltzmann constant $k_B = 1$.  
The analysis given in Sections II and III can be repeated 
for the case of finite temperature by implementing the 
following replacements\cite{OC}:
\begin{eqnarray}
\int d^4 x &\rightarrow& \int_x \equiv \int_{0}^{- i \beta} 
dx_0 \int d^3 x \nonumber \\
\Sigma \!\!\!\!\!\! \int d^4 p &\rightarrow& \Sigma 
\!\!\!\!\!\! \int_p \equiv \frac{ 2\pi}{-i \beta}
\sum_l \sum_k \int dp_2 \int dp_3  \\
\hat{\delta}^{(4)}(p-p') &\rightarrow& \hat{\delta}_T^{(4)} 
(p-p') \equiv \frac{-i \beta}{2 \pi} \delta_{ll'} 
\delta_{kk'} \delta (p_2 -p'_{2}) \delta (p_3 -p'_3) 
\nonumber
\end{eqnarray}
and for the photon propagator 
\begin{equation}
\int d^4 q \rightarrow \frac{2 \pi}{-i \beta} \sum_{l'} 
\int dq_1 \int dq_2 \int dq_3
\end{equation}

The gap equation now reads
\begin{equation}
1 \simeq \frac{2 \alpha}{\pi} T |eH| \int_{-\infty}^{\infty} 
dq_3 \int_0^\infty d\hat{q}_\perp^2 {\rm e}^{- \hat{q}_\perp^2} 
\sum_{l'} \frac{1}{Q_2 + 4 \pi^2 T^2 l'^2}~\frac{1}{Q_1 + 
\pi^2 T^2 (2 l' -1)^2} 
\label{tgap}
\end{equation}
where $Q_1 \equiv q_3^2 + m^2$ and $Q_2 \equiv q_3^2 
+ 2 |eH| \hat{q}_\perp^2$.  Following Ref. \cite{SDM}, we sum the 
series by means of the Poisson sum formula, which states 
that, if $c(\tau)$ is the Fourier transform of $b(\omega)$,
\begin{equation}
 c(\tau) = \int_{- \infty}^{\infty}  b(\omega) 
{\rm e}^{ -i \omega \tau}  d\omega,
\end{equation}
the following identity will hold:
\begin{equation}
 \sum_{l' = - \infty}^{\infty} b(l') = 
 \sum_{\lambda = - \infty}^{\infty} c( 2 \pi \lambda).
\end{equation}
Here, taking
\begin{equation}
   b(\omega) = \frac{1}{Q_2 + 4 \pi^2 T^2 \omega^2}~
\frac{1}{Q_1 + \pi^2 T^2 (2 \omega - 1)^2},
\end{equation}
we have
\begin{equation}
c(\tau) = \frac{1}{8 \pi T^2} \frac{1}{\sqrt{Q_1 Q_2}} 
\int_{-\infty}^{\infty} du \exp\left[-\left(i \frac{u}{2} 
+ \frac{|u|}{2 \pi T} \sqrt{Q_1}  
+ \frac{|\tau - u|}{2 \pi T} \sqrt{Q_2}\right)\right].
\end{equation}

After integrating over $u$, we make use of the Poisson sum 
formula to rewrite the summation in Eq.(\ref{tgap}) in 
terms of a more manageable series:
\begin{eqnarray}
&& \sum_{l'} \frac{1}{Q_2 + 4 \pi^2 T^2 l'^2} 
\frac{1}{Q_1 + \pi^2 T^2 ( 2 l' -1)^2}
\nonumber \\
 &~=~& \frac{1}{2T} \frac{1}{\sqrt{Q_1 Q_2}}
{\sum_{\lambda = 0}^{\infty}}' \left[\left({\rm e}^{\frac
{-\lambda \sqrt{Q_2}}{T}} + (-1)^\lambda {\rm e}^{-\frac
{\lambda \sqrt{Q_1}}{T}} \right) \frac{\sqrt{Q_1} + \sqrt{Q_2}}
{(\sqrt{Q_1} + \sqrt{Q_2})^2 + \pi^2 T^2} \right. 
\nonumber \\
 && ~~~~~~~~~~~~~~~~~~~~~~\left. \left(
{\rm e}^{\frac{-\lambda \sqrt{Q_2}}{T}} - (-1)^\lambda 
{\rm e}^{\frac{-\lambda \sqrt{Q_1}}{T}} \right) 
\frac{\sqrt{Q_1} - \sqrt{Q_2}}{(\sqrt{Q_1} - \sqrt{Q_2})^2 
+ \pi^2 T^2} \right], 
\label{finitt}
\end{eqnarray}
where the prime on the summation sign means that the 
$\lambda = 0$ term is counted with half weight.  
Subsituting Eq.(\ref{finitt}) into Eq.(\ref{tgap}) and 
summing over the infinite geometric series, we obtain
\begin{eqnarray}
 1 &~\simeq~& \frac{\alpha}{\pi} |eH| \int_{-\infty}
^\infty dq_3 \int_0^\infty d\hat{q}_\perp^2  
\frac{{\rm e}^{- \hat{q}_\perp^2}}{\sqrt{Q_1 Q_2}} 
\nonumber \\
 & & \cdot \left[\left(\frac{1}{1-{\rm e}^{-\frac
{\sqrt{Q_2}}{T}}} + \frac{1}{1+{\rm e}^{-\frac{\sqrt{Q_1}}
{T}}} - 1 \right) \frac{\sqrt{Q_1} + \sqrt{Q_2}}{(\sqrt{Q_1} 
+ \sqrt{Q_2})^2 + \pi^2 T^2} \right. \nonumber \\
 & & ~~+ \left. \left(\frac{1}{1-{\rm e}^{-\frac{\sqrt{Q_2}}
{T}}} - \frac{1}{1+{\rm e}^{-\frac{\sqrt{Q_1 }}{T}}} \right)  
\frac{\sqrt{Q_1} - \sqrt{Q_2}}{(\sqrt{Q_1} - \sqrt{Q_2})^2 
+ \pi^2 T^2} \right]. 
\label{gapt}
\end{eqnarray}
Note that the only approximations we have made so far are 
the quenched, ladder approximation (Eq.(\ref{massop})), 
the small $\hat{q}_\perp$ approximation, and keeping only 
the $k'' = 0$ terms in Eq.(\ref{SDparallel}).  Aside from 
these approximations, the finite-temperature gap equation, 
Eq.(\ref{gapt}), is exact in its dependence on the coupling 
constant, the magnetic field, and the temperature.  

We shall consider below the zero, low, and high temperature 
limits of Eq.(\ref{gapt}).  For $T = 0$, Eq.(\ref{gapt}) is 
reduced to
\begin{equation}
 1 \simeq  \frac{\alpha}{\pi} |eH| \int_{-\infty}^\infty 
dq_3 \int_0^\infty d\hat{q}_\perp^2 
\frac{{\rm e}^{-\hat{q}_\perp^2}}{\sqrt{Q_1 Q_2}} 
\frac{1}{\sqrt{Q_1} + \sqrt{Q_2}} 
\label{gap0}
\end{equation}
which, one can easily check, is just what Eq.(\ref{gapeq}) 
becomes when the integration there over $q_4$ is done.  
Thus, we have recovered the $T = 0$ result.    

For the low temperature case, we can approximate the sum 
over $l'$ in Eq.(\ref{tgap}) by the $\lambda = 0$ term 
(with half weight) on the right hand side of 
Eq.(\ref{finitt}), the other terms being exponentially 
small.  Treating the thermal effects as a perturbation, 
we write $ m^2(T) = m^2_0 + \delta m^2$ with 
$\delta m^2 << m_0^2$, where $m_0$ is the fermion 
dynamical mass for $T = 0$ (i.e., the solution in 
Eq.(\ref{dynamass})).  At small $T$, the difference 
between Eq.(\ref{tgap}) and Eq.(\ref{gap0}) yields
\begin{equation}
 \delta m^2 = - 2\pi^2 T^2 \frac{I_2}{I_1}
\end{equation}
where
\begin{eqnarray}
  I_1 &~=~& \int_{-\infty}^{\infty} dq_3 \int_{0}^{\infty} 
d\hat{q}_\perp^2 {\rm e}^{-\hat{q}_\perp^2} \frac{1}{Q}  
\frac{1}{\sqrt{Q_2}} \frac{1}{\sqrt{Q} + \sqrt{Q_2}} 
\left(\frac{1}{\sqrt{Q}} + \frac{1}{\sqrt{Q} + \sqrt{Q_2}} 
\right) \nonumber \\
 I_2 &~=~& \int_{-\infty}^{\infty} dq_3 \int_{0}^{\infty} 
d\hat{q}_\perp^2 {\rm e}^{-\hat{q}_\perp^2} 
\frac{1}{\sqrt{Q}} \frac{1}{\sqrt{Q_2}} \frac{1}{(\sqrt{Q} + 
\sqrt{Q_2})^3}   
\end{eqnarray}
with $Q \equiv  m_0^2 + q_3^2$.  Note that the infrared regions 
of $q_3$ and $\hat{q}_\perp^2$ dominate the integrals, just as 
in the $T = 0$ case.  More importantly, both $I_1$ and $I_2$ 
are finite and positive so that $\delta m^2$ is negative (and 
small, being proportional to $T^2$).  Thus, thermal effects 
tend to reduce the fermion dynamical mass, i.e., chiral symmetry 
tends to be restored as the temperature increases, as discussed 
earlier.  However, we should stress that, so long as $T$ is 
small, a non-zero dynamical mass will be generated in the 
presence of the magnetic field.   

At high temperatures ($T > \sqrt{|eH|}$), the $l' = 0$ term 
dominates the sum in Eq.(\ref{tgap}), thus yielding
\begin{equation}
 1 \simeq  \frac{2 \alpha}{\pi } T |eH| \int_{-\infty}^{\infty} 
 dq_3 \int_0^\infty d\hat{q}_\perp^2 {\rm e}^{- \hat{q}_\perp^2} 
 \frac{1}{Q_2} \frac{1}{Q_1 + \pi^2 T^2}. 
\end{equation}
Since the dominant contributions to the $q_3$-integral come from 
small values of $q_3^2$, we may approximate the denominator 
$(Q_1 + \pi^2 T^2)$ by $[m^2(T) + \pi^2 T^2]$.  After evaluating 
the integrals, we obtain
\begin{equation}
 m^2(T) + \pi^2 T^2 \simeq \alpha T \sqrt{2 \pi |eH|} 
\label{solu}
\end{equation}
Obviously, at weak couplings ($\alpha \ll 1$) and at high 
temperatures ($T > \sqrt{|eH|}$), there is no non-negative  
solution for $m^2(T)$, i.e., no chiral symmetry breaking solution.  
Thus, both the high temperature and the low temperature solutions 
indicate that chiral symmetry will be restored at high 
temperatures.

If we define the critical temperature for chiral symmetry breaking 
to be the temperature at which $m^2(T)$ vanishes, i.e., $m^2(T_c) 
= 0$, Eq.(\ref{solu}) provides an estimate of the critical 
temperature:
\begin{eqnarray}
  T_c &~\simeq~~& \frac{\alpha}{\pi^2} \sqrt {2 \pi |eH|}
\nonumber \\
  &~\sim~& 1.5 \times 10^{-2}~{\rm eV}~\sqrt{\frac{|H|}
  {10^4~{\rm gauss}}}.
\label{Tc}
\end{eqnarray}
For a magnetic field of any given strength, the critical 
temperature can be more exactly calculated by numerically solving 
the gap equation, Eq.(\ref{gapt}), with $Q_1$ replaced by $q_3^2$.

Finally, we consider the suggestion by Gusynin {\it et al.}\cite{GMS} 
that the chiral symmetry breaking solution found above may play 
a role during the electroweak phase transition in the early universe.  
The electroweak phase transition took place at a temperature of order 
100 GeV.  From Eq.(\ref{Tc}), this requires a magnetic field of 
$10^{30}$ gauss or stronger, which is significantly larger than any 
estimates of the magnetic field strength at the time of the 
electroweak phase transition\cite{HEW}.  We therefore conclude that 
the chiral symmetry breaking solution considered here does not play 
any role in the electroweak phase transition.

\newpage
\section{Conclusion}

We have described a formalism for studying the physics of chiral 
symmetry breaking in an external field via the nonperturbative 
Schwinger-Dyson equation.  The $E_p$-representation for the 
fermion propagator proposed here has the advantage that the 
dependence on the operator $\Pi_\mu$ is removed.  It also has 
the advantage over Schwinger's proper time formalism for calculating 
finite-temperature effects.  

We have applied our method to examine chiral symmetry breaking in 
QED in a constant external magnetic field.  We find that, even when 
the coupling constant is small, an external magnetic field can trigger 
the dynamical breaking of chiral symmetry in QED, with the dynamical 
mass of the fermion given by Eq.(\ref{dynamass}).  Our result agrees 
with that found in Ref. \cite{GMS} which used an approach rather 
different from ours.  The existence of the Nambu-Goldstone boson 
of chiral symmetry breaking is demonstrated in the Appendix by showing 
that the same solution (Eq.(\ref{dynamass})) solves the Bethe-Salpeter 
equation for the bound-state Nambu-Goldstone boson.

We have also obtained an estimate of the critical temperature $T_c$ 
of the aforementioned chiral symmetry breaking, Eq.(\ref{Tc}).  Chiral 
symmetry is a good symmetry above this critical temperature, and it is 
spontaneously broken at temperatures below $T_c$.  Our result renders 
invalid the suggestion in the literature\cite{GMS} that the chiral 
symmetry breaking solution described above may be relevant for the 
electroweak phase transition in the early universe.  

There remain several interesting questions which we are investigating.  
For instance, how do other background field configurations affect 
chiral symmetry breaking?  (As an example, it is expected that an 
electric field would tend to break up the condensate and destabilize 
the vacuum, thus inhibiting chiral symmetry breaking\cite{DW}.)  Are 
there strong-coupling solutions of chiral symmetry breaking in an 
external magnetic field?  What are the effects of additional 
four-fermion interactions?  Recall that four-fermion operators play 
a crucial role in obtaining a consistent chiral symmetry breaking 
solution in quenched, ladder QED\cite{BLL}.  We hope to report our 
findings in the near future.

\newpage
\begin{center} 
{\bf ACKNOWLEDGEMENTS}\\
\end{center}

This work was supported in part by the U.S. Department of Energy 
under Grant No. DE-FG02-84ER40163 and DE-FG05-85ER-40219 Task A.
Part of this work was done some time ago when C.N.L. was visiting 
UNC and when Y.J.N. was visiting M.I.T. and the Institute for Advanced 
Study.  They thank the respective faculties for their hospitality.  
\raggedbottom

\newpage
\appendix
\section{\bf Nambu-Goldstone Bosons}

We verify in this appendix that the dynamical mass found in 
Eq.(\ref{dynamass}) corresponds to a spontaneous chiral symmetry 
breaking solution by examining the Bethe-Salpeter equation of the 
Nambu-Goldstone (NG) boson.  In the quenched, ladder approximation,
the Bethe-Salpeter equation of the NG boson has the form\cite{Mir}

\begin{equation}
\phi(x, y; P) = -~i e^2 \int d^4x' \int d^4y' G_A(x, x') 
\gamma^\mu \phi(x', y'; P) \gamma^\nu G_A(y', y) 
D_{\mu \nu}(y'-x').
\label{BSx}
\end{equation}
In the $E_p$-representation, this can be expressed as

\begin{eqnarray}
(2 \pi)^4 \hat{\delta}^{(4)}(p-p') \tilde{\phi}(p; P) 
&=& 
-~i e^2 \int d^4x d^4y \Sigma \!\!\!\!\!\! \int \frac{d^4p''}
{(2\pi)^4} \frac{1}{\gamma \cdot \bar{p} + m} 
\bar{E}_p(x) \gamma^\mu E_{p''}(x) \nonumber \\
& &
\cdot~ \tilde{\phi}(p''; P) \bar{E}_{p''}(y) \gamma^\nu 
E_{p'}(y) \frac{1}{\gamma \cdot \bar{p}' + m} D_{\mu\nu}(y-x), 
\label{BSp}
\end{eqnarray}
where 
\begin{equation}
\phi(x, y; P) = \Sigma \!\!\!\!\!\! \int \frac{d^4p}{(2 \pi)^4} 
E_p(x) \tilde{\phi}(p; P) \bar{E}_p(y)
\end{equation}
and $m$ is the dynamically generated fermion mass.

If we introduce
\begin{equation}
\chi(p; P) \equiv \left(\gamma \cdot \bar{p} + m \right) 
\tilde{\phi}(p; P) \left(\gamma \cdot \bar{p} + m \right),
\label{defchi}
\end{equation}
we find 

\begin{eqnarray}
(2 \pi)^4 \hat{\delta}^{(4)}(p-p') \chi(p; P) 
&=& 
-~i e^2 \int d^4x d^4y \Sigma \!\!\!\!\!\! \int \frac{d^4p''}
{(2\pi)^4} 
\bar{E}_p(x) \gamma^\mu E_{p''}(x) \nonumber \\
& &
\cdot~ \frac{1}{\gamma \cdot \bar{p}'' + m} \chi(p''; P) 
\frac{1}{\gamma \cdot \bar{p}'' + m} \nonumber \\
& &
\cdot~ \bar{E}_{p''}(y) \gamma^\nu E_{p'}(y) D_{\mu\nu}(y-x). 
\label{BSchi}
\end{eqnarray}
Note the similarity of this equation to Eq.(\ref{SDfull}).  
After integrating over $x$ and $y$ (see Eq.(\ref{xintfinal}) 
and Eq.(\ref{x'intfinal})) as well as over $p_0''$, $p_2''$ 
and $p_3''$, we obtain (in the Feynman gauge) 

\begin{eqnarray}
\chi(p; P) \delta_{kk'} 
&~=~&
-~i e^2 (2 |eH|) \sum_{k''} \sum_{\{\sigma\}} \int \frac
{d^4\hat{q}}{(2 \pi)^4} 
\frac{{\rm e}^{-i {\rm sgn}(eH)(n-n''+\tilde{n}''-n')\varphi}}
{\sqrt{n!n'!n''!\tilde{n}''!}} 
\nonumber \\
& &
\cdot~ \frac{{\rm e}^{- \hat{q}_\perp^2}}{\hat{q}^2} 
J_{nn''}(\hat{q}_\perp) J_{\tilde{n}''n'}(\hat{q}_\perp)  
\Delta \gamma^\mu \Delta'' \nonumber \\
& &
\cdot~ \frac{1}{\gamma \cdot \bar{p}'' + m} \chi(p''; P) 
\frac{1}{\gamma \cdot \bar{p}'' + m} 
\tilde{\Delta}'' \gamma_\mu \Delta',
\label{SDchi}
\end{eqnarray}
where $\bar{p}''_0 = p_0 + q_0$, $\bar{p}''_1 = 0$, 
$\bar{p}''_2 = -~{\rm sgn}(eH) \sqrt{2|eH|k''}$, 
$\bar{p}''_3 = p_3 + q_3$.  

We are interested in the $P = 0$ behavior of $\chi(p; P)$, which is 
expected to have the form
\begin{equation}
\chi(p; 0) = A(p) \gamma_5,
\end{equation}
where $A(p)$ is a scalar function.  It follows from Eq.(\ref{SDchi}) 
that 
\begin{eqnarray}
A(p) \delta_{kk'} 
&~=~&
i e^2 (2 |eH|) \sum_{k''} \sum_{\{\sigma\}} \int \frac
{d^4\hat{q}}{(2 \pi)^4} 
\frac{{\rm e}^{-i {\rm sgn}(eH)(n-n''+\tilde{n}''-n')\varphi}}
{\sqrt{n!n'!n''!\tilde{n}''!}} 
\nonumber \\
& &
\cdot~ \frac{{\rm e}^{- \hat{q}_\perp^2}}{\hat{q}^2} 
J_{nn''}(\hat{q}_\perp) J_{\tilde{n}''n'}(\hat{q}_\perp)  
\frac{G_1 A(p'')}{\bar{p}''^2 + m^2},
\label{SDA}
\end{eqnarray}
with $G_1$ defined in Eq.(\ref{GWdef}).  After taking the small 
$\hat{q}_\perp$ approximation, Eq.(\ref{Japprox}), this is 
simplified to 
\begin{equation}
A(p) \delta_{kk'} = 
i e^2 (2 |eH|) \sum_{k''} \sum_{\{\sigma\}} \delta_{nn''} 
\delta_{\tilde{n}''n'} \int \frac{d^4\hat{q}}{(2 \pi)^4} 
\frac{{\rm e}^{- \hat{q}_\perp^2}}{\hat{q}^2} 
\frac{G_1 A(p'')}{\bar{p}''^2 + m^2}.
\end{equation}

If we consider the infrared behavior of $A(p)$, we have
\begin{equation}
A(0) \simeq 
-~i e^2 (4 |eH|) \int \frac{d^4\hat{q}}{(2 \pi)^4} 
\frac{{\rm e}^{- \hat{q}_\perp^2}}{\hat{q}^2} 
\frac{A(q_\parallel)}{2 |eH| \hat{q}_\parallel^2 + m^2},
\end{equation}
where we have performed the spin summation and ignored the 
$k'' = 1$ term, in accordance with the approximation used in 
obtaining Eq.(\ref{SDsigma0}).  Transforming to Euclidean 
space and noting that the integral is dominated by contributions 
from small $\hat{q}_\parallel^2$ so that $A(q_\parallel)$ in 
the integrand can be approximated by $A(0)$, we secure 
\begin{equation}
1 \simeq e^2 (4 |eH|) \int \frac{d^4\hat{q}}{(2 \pi)^4} 
\frac{{\rm e}^{- \hat{q}_\perp^2}}{\hat{q}^2} 
\frac{1}{2 |eH| \hat{q}_\parallel^2 + m^2},
\end{equation}
which is the same as the gap equation, Eq.(\ref{gapeq}).  Thus, 
the dynamical mass found in Eq.(\ref{dynamass}) provides a 
consistent solution to the Bethe-Salpeter equation for the NG boson.


\begin{references}
\bibitem{OC}
The orthonormality and the completeness relations are 
replaced respectively by $\displaystyle{\int}_{x} 
\bar{E}_{p'}(x) E_p(x) = (2 \pi)^4 \hat{\delta}_T^{(4)}(p-p')$ 
and $\Sigma \!\!\!\!\! \displaystyle{\int}_{p} E_p(x) \bar{E}_p(y) 
= (2 \pi)^4 \delta^{(4)}(x-y)$.
\end{references}
\end{document}